# Analysis of Cosmological Generalized Reduced Void Probability Functions Constrained by Observations and Numerical Simulations

Keith Andrew, David Barnaby, Lisa Taylor
*Department of Physics and Astronomy*
*Western Kentucky University*
*Bowling Green, KY 42101, USA*

Using survey data and numerical ΛCDM modeling we establish an optimized fit to the generalized Reduced Void Probability Function, RVPF, of Mikjian used to establish a statistical foundation to any physical process associated with hierarchical clustering. We use a numerical N-body cosmological simulation code, GADGET-2, to investigate the sensitivity of the distribution of voids characterized by the RVPF to a general hierarchical reduced void parameter, *a*. The void parameter is related to the Levy stability index of the distribution, $\alpha = 1- a$, and Fischer critical exponent, $\tau = 2 - a$, used in clustering models. We numerically simulate the evolution of the universe with $\Omega=1$ from a redshift of z=50 to the current epoch at z=0 in order to generate RVPFs. GADGET-2 is an N-body/smoothed particle hydrodynamics, SPH, code that we ran in MPI parallelizable mode on an HPC Beowulf cluster. The numerical data sets are compared to observational data from the Sloan digital sky Survey, SDSS, CfA, the Deep2 Galaxy Redshift Survey, and the 2dF Survey. We find the best value of the parameter *a* occurs near the Negative Binomial reduced void probability function but exhibits a departure from perfect scaling over the z values studied.

## I. INTRODUCTION

The large scale structure and distribution of matter in the universe is formed by clusters, filaments, sheets, and a number of void structures[1] making up a cosmic web that is indicative of the underlying physical laws.[2] The role of voids as basic ingredients of large scale cosmic structure is now well established, but the definition of what exactly constitutes a void is still evolving.[3,4] At first, voids were described as large regions devoid of galaxies, but the current view is more intricate than emptiness.[5,6] As the overall statistics of cosmic voids improves,[7] and as better observational data about the mass density enclosed in void structures is acquired, the cosmic void contribution to understanding the large scale structure of the universe is providing strong constraints on the nature of cosmic evolution. One reason for this is that galaxy groups located in void regions are relatively rare and there abundance can serve as a sensitive probe of non-Gaussianity.[8] In addition more thorough observational data can be coupled with focused numerical modeling based on scaling laws to yield a detailed and more dynamic view of void structure and evolution.[9,10] Several methods for finding voids have been refined and compared:[11] void numbers can be constrained by nonlinear galactic clustering,[12] by examining void galaxy correlations,[13] using the two point correlations between density perturbations,[14] counting the large galaxy distributions around voids,[15] and using the nearby profiles of the overall galactic velocity field.[16]

The increasing amount of observational survey data on voids has given rise to a number of efforts to analytically and numerically model voids, their abundance, essential shapes and overall distribution in space and time.[17][18] Generalized statistical models with hierarchical clustering can be implemented on all scales[19] and compared to observed space density profiles.[20] Special function sets, such as flaglets, make for increased efficiency and higher precision modeling.[21] Scaling features of voids and their connection to fractal models have been studied in detail by Gaite.[22][23] Studies of the scaling of moments of count distributions with characteristic cell sizes have been proposed as a method of obtaining the scaling dimension of the cosmic system and the proposed matching scaling laws.[24][25] Voids have been used as a test of dark matter models[26] and to pioneer advanced statistical methods.[27] A discussion of the use of the three point function instead of the more common two point methods is pursued in Takada and Jain.[28] A recent local search for low density structures that are void like in many ways restricted to the local supercluster out to 40 Mpc resulted in a count of 89 voids with sizes from 12 to 24 Mpc and absolute magnitudes ≤18.4.[29] In addition several numerical studies[30] have been done for standard LCDM cosmology[31] and de Sitter models,[32] some including halo effects.[33] Work by Hamaus indicates that using void-galaxy correlations one could determine the absolute galaxy bias in a survey with wide ranging implications to test general relativity on a large scale.[34] Robust void statistics can also be used to connect to and trace dark energy.[35] There are also some anomalous observations that enter the field of cosmic voids such as the analysis of the CMB Cold Spot[36] and the large void structure found in Bootes.[37] There are a growing number of robust data sets that have been used to compare void sampling and modeling, principally the CfA slice[38], Deep 2 Galaxy Redshift Survey,[39] the 2dF Survey,[40][41][42] and the expanding SDSS dataset.[43][44]

In this paper we will use the combined data sets: CfA, 2dF, Deep 2, and SDSS with a numerical study using a cosmic numerical simulation code[45], GADGET-2[46] to analyze void probability functions,[47] VPFs, and reduced void probability functions,[48] RVPFs, to determine the best fit to a generalized reduced void probability function. The remainder of this work is organized as follows: in Section 2 we define the generic void probability functions and reduced void probability functions that are used in hierarchical scaling models, of special interest here is the statistically motivated generalized RVPFs linked by a common grand canonical ensemble through a single parameter, Section 3 introduces the GADGET-2 generated data and the collected survey data from CdF, 2Df, Deep 2, and SDSS to construct the RVPFs of interest and a match to the generalized RVPF, match to void density is used to examine the z dependence of a the void model, finally concluding remarks are made in Section 4.

## 2. VOID PROBABILITY FUNCTIONS

As a brief overview of VPFs and the generalized statistical VPF model we follow the conventions of Mekjian[49] and Conroy[50] to define the void probability function and to relate it to a scaling model. The VPF is defined as the probability of finding no galaxies inside a sphere or radius R, randomly placed within a sample. For spherical volume elements the VPF can be expressed as:

$$P_o(R) = \exp\left[\sum_{p=1}^{\infty} \frac{-\overline{N}(R)^p}{p!} \overline{\xi}_p(R)\right] \quad (1)$$

where R is the sphere radius, $\overline{N}$ is the average number of galaxies in the sphere, and $\overline{\xi}$ is the volume averaged p-point correlation function defined as

$$\overline{\xi}_p = \frac{\int \xi_p dV}{\int dV}. \quad (2)$$

Here we will apply the hierarchical ansatz to relate all higher order correlation functions to the two point function, $\xi_2 = \xi$, by

$$\overline{\xi}_p = S_p \overline{\xi}_2^{p-1} = S_p \overline{\xi}^{p-1}, \quad p \geq 3 \quad (3)$$

for scaling coefficients $S_p$ giving a simplified VPF as:

$$P_o = \exp\left[-\sum_{p=1}^{\infty} \frac{(-\overline{N})^p}{p!} S_p \overline{\xi}^{p-1}\right]. \quad (4)$$

To isolate the effects of the scaling coefficients Fry[51] introduced the reduced void probability distribution, $\chi$, RVPD, with an independent variable $x = \overline{N}\overline{\xi}$:

$$\chi(x) = -\ln(P_o)/\overline{N} = \sum_{p=1}^{\infty} \frac{S_p}{p!}(x)^{p-1}. \quad (5)$$

Some simple void distributions can be effectively understood by choosing an analytical model that allows for the use of the scaling factors. One set of scaling factors that appear in several galaxy counts corresponds to $S_p = (p-1)!$ then the RVPF and the fluctuations, using brackets to clearly denote the average, are

$$\chi = \sum_{p=1}^{\infty}\left(\frac{1}{p}\right)x^{p-1} = \frac{\ln(1+x)}{x}$$
$$\delta N^2 = \langle N^2 \rangle - \langle N \rangle^2 = \langle N \rangle - \xi\langle N \rangle^2 \quad (6)$$

which is the phenomenological Negative Binomial distribution. Several other analytical cases are collected in the Table 1 and have served as potential distributions for void clustering, in particular, using the parameter "*a*" introduced by Mekjian[52] to describe a general VPF/RVPF noting that when *a*=1/2 this corresponds to the thermodynamic model that was developed by Saslaw and Hamilton[53] and that when *a*=1 we have the negative binomial model that was investigated by Carruthers, Duong-van and Croton.[54][55][56] Croton had the best overall fit with the negative binomial and noticed small departures for large x and considered a Gaussian model, a thermodynamic model, a lognormal distribution, and a model based on a BBGKY distribution. In Mekjian's statistical construction the parameter *a* is related to the Levy stable index α and the Fischer critical exponent τ.

The negative binomial and thermodynamic models are special cases of a general statistical Hierarchical Scaling Model developed by Mekjian. He introduces a generating function for a probability distribution that can be expressed in terms of the combinants $C_k$ and then the VPF and RVPF can be expressed in terms of the grand canonical partition function as

$$\sum_{n=0}^{\infty} P_n u^n = \exp\left(\sum_{k=1}^{\infty} C_k \left(u^k - 1\right)\right)$$
$$Poisson: C_k = C_1 \delta_{1k}$$
$$Negative-Binomial: C_k = \frac{xt^k}{k} \quad Z_{gc} = (1-tu)^{-x} \quad (7)$$
$$VPF: P_0 = \frac{1}{Z_{gc}} \quad RVPF: \chi = \frac{-\ln P_0}{\langle N \rangle} = \frac{\ln Z_{gc}}{\langle N \rangle}$$

where the Poisson and Negative Binomial distributions have explicit analytical forms. The general grand canonical partition function can be expressed in terms of a new parameter *a* using the hypergeometric function, in terms of Pochhammer symbols as

$$\ln(Z_{gc}) = (xtu) \,_2F_1(a;1;2;tu) = \frac{x\left(1-(1-t)^{1-a}\right)}{1-a}$$
$$_2F_1(a;b;c;z) = \frac{[a]_m [b]_m}{[c]_m m!} z^m \quad (8)$$
$$[a]_m = a(a+1)(a+2)....(a+m-1) = \frac{\Gamma(a+m)}{\Gamma(a)}$$

For *a*=1 we get the negative binomial and for *a*=1/2 this is the thermodynamic model. This model is not the same as the Holtsmark distribution which is a stable α=3/2 distribution that can be expressed as a sum of hypergeometric functions and is used in astrophysical applications.

| Model | $S_p$ | RVPF $\chi$ |
|---|---|---|
| Gaussian Model | Only p=2 | $\chi = 1 - x/2$ |
| Minimal Model | $S_p$=1 for all p | $\chi = (1 + e^{-x})/2$ |
| Thermodynamic Model | $S_p$=(2p-3)!! | $\chi = \left((1+2x)^{1/2} - 1\right)/(x)$ |
| Negative Binomial Model | $S_p$=(p-1)! | $\chi = \ln(1+x)/(x)$ |
| Generalized Hierarchical | Generalized $S_p$ $S_p = \dfrac{\Gamma(p+a-1)}{\Gamma(a)\,a^{p-1}}$ | $\chi_a(x) = \dfrac{1}{(1-a)\left(\dfrac{x}{a}\right)}\left[\left(1+\dfrac{x}{a}\right)^{1-a} - 1\right]$ |

**Table 1** Sample analytic and model Reduced Void Probability Functions as they apply to different scaling coefficient models, notice that in terms of the Generalized Hierarchical Model that *a*=1 yields the negative binomial model and *a*=1/2 yields the thermodynamic model.

For $1 \geq a \geq 0$ the Levy index α=1-a, where the Levy index is used to characterize the behavior of non-Gaussian probability distributions that exhibit asymptotic power law forms and the Fischer critical exponent is τ = (2 – a), where the Fischer critical exponent describes the power law fall off of clusters of matter or voids at a phase transition point. Levy stable distributions have a Levy index 0 < α ≤ 2 and for a Levy index strictly less than 2 the probability distribution exhibits an asymptotic power law ~ $|x|^{-1-\alpha}$ with a heavy tail. We will use the observational data and numerical runs of GADGET-2 to estimate the value of *a* in the generalized reduced void probability function and therefore the exponent of the asymptotic void power law.

## II. VPF Structure in GADGET-2

We use the N-body code GADGET-2[57] to numerically investigate the distribution and evolution of voids in the universe. GADGET-2 is an N-body/smoothed particle hydrodynamics, SPH, code available from Springer Vogel that we ran in MPI parallelizable mode on a Beowulf cluster. GADGET-2 employs a tree method to calculate gravitational forces. Optionally, the code uses a tree-PM algorithm based on an explicit split in Fourier space between long-range and short-range forces. This combination provides high performance while still retaining the full spatial adaptivity of the tree algorithm. By default, GADGET-2 expands the tree multipoles only to monopole order, in favor of compact tree storage, a cache-optimized tree-walk, and consistent and efficient dynamic tree updates. The cell-opening criterion used in the tree walk is based on an estimator for the relative force error introduced by a given particle-cell interaction, such that the tree force is accurate up to a prescribed maximum relative force error. The latter can be lowered arbitrarily, if desired, at the expense of higher calculation times.

The particle-mesh, PM, part of GADGET-2 solves Poisson's equation on a mesh with standard fast Fourier transforms, based on a counts in cells mass assignment and a four-point finite differencing scheme to compute the gravitational forces from the potential. Here we consider simulations of N = 2 x $10^6$ galaxies with a mass of $10^{10}$ solar masses on a cube of length 350 Mpc/h , Ω=1, a dark energy content of 0.727, Hubble parameter h=0.70, spatial curvature k=0, starting at z=50 and evolving to z=0 in commoving coordinates with periodic boundary conditions. We also vared the z=50 over density as an initial condition from zero to 0.01 ppb to 100 ppb and track the evolution of the VPFs and RVPFs as spheres. Fig. 1 shows a typical initial and final volume set for an over density of 1ppb expressed as the density contrast function: $\delta = (\rho - \bar{\rho})/\bar{\rho}$.

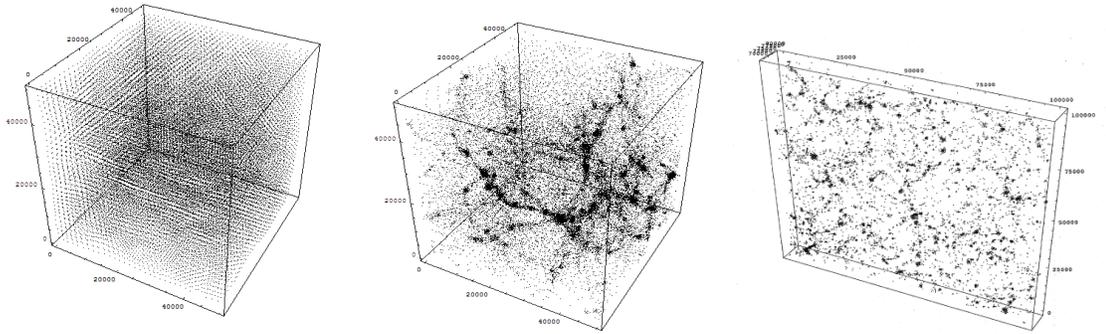

Fig. 1 Two cubical volumes 350/h Mpc on edge for $10^6$ galaxies of $10^{10}$ solar masses starting a z=0 to z=50 with an over density of 1ppb at z=50. *Left*: initial condition set at z=50, *Center*: image of z=3 distribution with the emergence of structure, notice that although there are many under-dense regions compared to the initial configuration there are not many large regions that are actually empty, *Right*: a 50/h Mpc thin slice at z=0 showing nonuniform structure formation and the emergence of voids where the under-density is very close to zero in many areas. The Larger voids do emerge as low density structures that evolve to still lower density but nearby streaming begins to contribute to void formation.

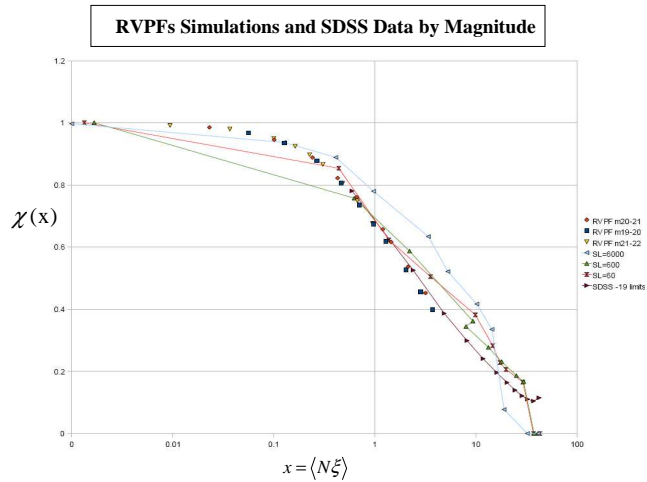

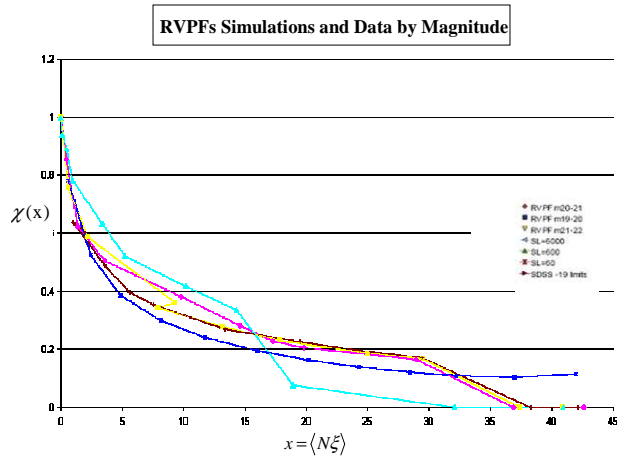

Fig. 2 *Top:* Comparison of Reduced Void Probability Functions vs. x in a semi-log plot for luminosity bins for SDSS mag. < 19, combined 2dF and SDSS mag. 19-20, mag. 20-21 and mag. 21-22, and GADGET-2 at softening lengths of 60, 600 and 6000. *Bottom*: Comparison plots for Reduced Void Probability functions as on the left vs. x.

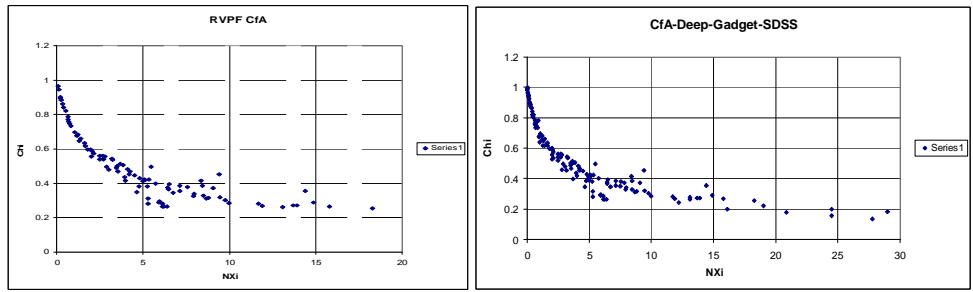

Fig. 3 Scatter Plots of Reduced Void Probability survey data. *Left*: the CfA dataset which is very sparse for x> 10, *Right* the Deep CfA, SDSS and GADGET-2 data sets which provide data out to nearly x= 30.

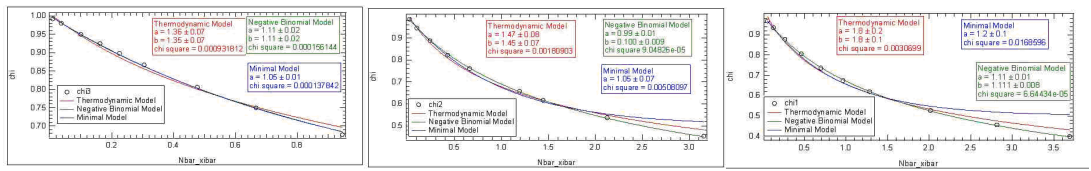

Fig. 4 Fit to data for Reduced Void Probability Functions for the Thermodynamic, Negative Binomial and Minimal Model for different ranges of x. *Left:* all three models display a good match near the origin. *Center:* when extended out to larger regions the three models deviate

with the Minimal model predicting higher values. *Right:* At the largest values where we have survey data is the Negative Binomial model results in the best overall description of the Reduced Void Probability Function.

Over all data sets and simulations treating *a* as a free parameter the optimal RVPF fit value is $a=0.89 \pm 0.03$, however the void size and relative density does not scale uniformly in the simulations but depends upon z. We do not observe a change in void density that is simply tracking the cosmic expansion. The bulk mass motions distort the reduced void probability functions which in part can be seen as a z dependence of the original shape functions and can be tracked as a function of void density. To check on the extent of the changes we follow the model of Ricciardelli[58][59] where we match the void density profile and z dependence for our total data set using a radial density profile function; we select four spherical shells around each void for a set of voids at different R and threshold densities for a fixed z then we increment z in steps of five and match the linear coefficients in the exponents of the density profile. Typical voids continue to empty as they evolve, from the linear fits in redshift to the exponents α and β expressed as:

$$\frac{\rho}{\rho_c} = \left(\frac{r}{R_c}\right)^\alpha \exp\left(\left(\frac{r}{R_c}\right)^\beta - 1\right) \qquad (7)$$

we find that:

$$\alpha = (0.09 - 0.14z + 0.003z^2) \pm 0.01$$
$$\beta = (1.19 - 0.02z - 0.0008z^2) \pm 0.04 \qquad (8)$$

for values of z from 0 to 15 in GADGET-2 where at z=50 the void density is the same as average cosmic density from our initial conditions and the error values represent the average standard error for each coefficient indicating the difficulty we have in arriving at significant quadratic terms. The older voids are not as spherical in overall shape and typically have lower densities as matter continues to exit the void volume leading to a sharper void edge feature. For our simulations the largest effects are observed to be centered on changes in α, the scaling of the size.

## IV Conclusions

In this work we have modeled cosmic voids as spheres and ignored their overlap, we found the generalized reduced void probability function has an optimal fit near the Negative Binomial value of $a = 0.89 \pm 0.03$, giving a Levy stable index of $\alpha = 0.09$ and a Fischer critical exponent of $\tau = 1.09$ but that for large x the void shapes deviate from the hierarchical scaling given by the z dependence in the void densities. Numerical algorithms in general are not specifically designed to model voids; in adaptive routines it is the empty regions that may not be treated in sufficient detail to prevent systematic errors. A way to overcome this is to have the adaptive focus on the void interior and then remove the need to analyze only spherical shapes by treating arbitrary shapes; both of these issues have been handled by Ricciadelli et. al. Improved analysis and collection of survey data in both scale and resolution is also improving the overall statistical void analysis[60] outlook thereby providing constraints on models. Such observations, when coupled with increases in calculation speed and available memory needed for computational simulation techniques,[61] will lead to considerable refinement and growth in cosmological void applications.

## ACKNOWLEDGMENTS

We gratefully acknowledge fellowship support for Lisa Taylor and from the NASA KY Space Grant Consortium, this work was made possible by two NASA KSGC awards that provided financial and facilities support at the Institute for Astrophysics and Space Science and HPC computer usage and the GADGET-2 team headed by Volker Springel for providing code and support at: http://www.mpa-garching.mpg.de/galform/gadget/ .